# Slideflow: Deep Learning for Digital Histopathology with Real-Time Whole-Slide Visualization

James M. Dolezal[1]*, Sara Kochanny[1], Emma Dyer[1], Andrew Srisuwananukorn[2], Matteo Sacco[1], Frederick M. Howard[1], Anran Li[1], Prajval Mohan[3], Alexander T. Pearson[1]

[1]Section of Hematology/Oncology, Department of Medicine, University of Chicago Medical Center, Chicago, IL, USA
[2]Tisch Cancer Institute, Icahn School of Medicine at Mount Sinai, New York, NY, USA
[3]Department of Computer Science, University of Chicago, Chicago, IL, USA

*Correspondence to: james.dolezal@uchicagomedicine.org

## Abstract

Deep learning methods have emerged as powerful tools for analyzing histopathological images, but current methods are often specialized for specific domains and software environments, and few open-source options exist for deploying models in an interactive interface. Experimenting with different deep learning approaches typically requires switching software libraries and reprocessing data, reducing the feasibility and practicality of experimenting with new architectures. We developed a flexible deep learning library for histopathology called Slideflow, a package which supports a broad array of deep learning methods for digital pathology and includes a fast whole-slide interface for deploying trained models. Slideflow includes unique tools for whole-slide image data processing, efficient stain normalization and augmentation, weakly-supervised whole-slide classification, uncertainty quantification, feature generation, feature space analysis, and explainability. Whole-slide image processing is highly optimized, enabling whole-slide tile extraction at 40X magnification in 2.5 seconds per slide. The framework-agnostic data processing pipeline enables rapid experimentation with new methods built with either Tensorflow or PyTorch, and the graphical user interface supports real-time visualization of slides, predictions, heatmaps, and feature space characteristics on a variety of hardware devices, including ARM-based devices such as the Raspberry Pi.

# 1 Background

Histopathology slides of patient tissue and tumor specimens serve many purposes and are increasingly being captured and stored in digital formats. The advent of deep learning models for analyzing digital histopathology images has unlocked a new dimension from which we can extract clinically meaningful information [1]. These models not only accelerate and enhance pathology clinical workflows but also detect subtle morphological features that escape the human eye, improving diagnostic efficiency and accuracy [2], [3]. Furthermore, deep learning models allow genomic subtype classification directly from digital histopathology images [4]–[7], and they show promise as tools for patient risk stratification [8], prognosis [9], and treatment selection [10]. Digital histopathology may also increase accessibility of hematoxylin and eosin (H&E) and immunohistochemistry staining in low-resource settings through virtual staining, and these tools offer the ability to boost clinical workflow efficiency and provide advanced diagnostics that may otherwise be unattainable due to a limited number of trained pathologists or their absence altogether [11]. Despite the far-reaching potential of deep learning applications in digital histopathology, development complexity and access to computational resources remain barriers to widespread adoption. There is a growing need for accessible and efficient open-source software that functions as a platform to perform these analyses.

Efficient software design is crucial for deep learning research toolkits in digital histopathology. Model performance improves with the amount of training data provided [12], but data storage requirements may impose practical limitations on dataset sizes and experimental scope. Unified and efficient data storage can reduce data redundancy and optimize the computational resources needed for training models. Computationally efficient software can reduce data processing and storage requirements while shortening training time, enabling groups with varying equipment capabilities to train their own models or access pre-trained models for novel research objectives.

Additionally, it is vital for deep learning models that aim to support clinical decision-making to be transparent and interpretable [13]. Explaining how a model has reached its decision and the level of certainty associated with a prediction can help build clinician trust, which may help foster greater adoption of these tools into clinical practice. Software that seamlessly integrates explainability and uncertainty quantification presents a significant advantage in promoting the potential clinical utility of these deep learning tools.

As computational tools continue to gain importance in biological sciences, it is essential to ensure their accessibility to researchers with diverse computational backgrounds. This can be achieved through comprehensive documentation, intuitive code design, and active project development. When creating analytical tools for clinical applications, it is also important to consider that the target end-users may not possess extensive computational expertise. A graphical user interface (GUI) can offer an accessible entry point for such users, facilitating the deployment of deep learning tools in clinical settings.

In this work, we introduce Slideflow, a Python package that delivers an end-to-end suite of user-friendly tools for building, testing, explaining, and deploying deep learning models for histopathology applications. Slideflow aims to overcome challenges in computational efficiency, accessibility, and interpretability while empowering researchers and clinicians to harness the full potential of deep learning in digital histopathology.

## 2 Implementation

### 2.1 Technical overview

Slideflow is a Python package providing a library of deep learning tools implemented with both PyTorch and Tensorflow backends. It has been developed to provide an end-to-end toolkit for building and deploying deep learning histopathology applications for scientific research, including efficient data processing, model training, evaluation, uncertainty quantification, explainability, and model deployment in a graphical user interface (GUI). Slideflow includes a whole-slide user interface, Slideflow Studio, for generating predictions and heatmaps for whole-slide images in real time. Slideflow is easy to deploy, with distributions available on the Python Packaging Index (PyPI) and pre-built docker containers on Docker Hub.

### 2.2 Whole-slide image processing

Slideflow supports nine slide scanner vendors (Table 1) and includes two slide reading backends – cuCIM [14], an efficient, GPU-accelerated slide reading framework for TIFF and SVS slides, and VIPS [15], an OpenSlide-based framework which adds support for additional slide formats. The first step in processing whole-slide images (WSI) for downstream deep learning applications is slide-level masking and filtering, a process that determines which areas of the slide are relevant and which areas should be ignored. Slides can be manually annotated with Regions of Interest (ROIs) using the provided whole-slide visualization tool, Slideflow Studio, or using alternative programs such as QuPath [16] or Aperio ImageScope. Tissue detection can be performed with Otsu's thresholding, which masks areas of background. Pen marks and out-of-focus areas can be masked with Gaussian blur filtering.[*] Additionally, Slideflow includes an API for custom slide-level masking, with an example demonstrating how to use this API to apply a DeepFocus [17] model for detection of out-of-focus regions.

| Vendor | File format |
| --- | --- |
| Aperio | SVS |
| Philips | TIFF |
| Mirax | MRXS |
| Hamamatsu | NDPI |
| Leica | SCN |
| Ventana | BIF, TIF |
| Trestle | TIF |
| Sakura | SVSLIDE |
| Olympus | TIFF |

**Table 1. Slide scanner compatibility.** Images from Philips slide scanners must be exported from the scanner in TIFF format, as iSyntax files are currently unsupported.

After slide-level masking and filtering, WSIs are tiled into smaller sub-regions. A schematic of data flow during tile extraction is shown in **Fig. 1**. Image tiles can be buffered into binary TFRecord format, an efficient storage format that improves dataset iteration speed compared to iterating directly from WSIs. Image data can be encoded using either JPEG (lossy) or PNG (lossless) compression. Image tiles are extracted in a grid, with optional overlap or jitter for data augmentation. During tile extraction, images tiles can undergo an additional filtering step using either grayspace or whitespace filtering, potentially identifying tiles with high background content that were not successfully removed by Otsu's thresholding. Grayspace filtering is performed by converting images into the hue, saturation, value (HSV) spectrum, classifying pixels with saturation below a given threshold[*] as gray, and discarding the tile as background if the fraction of pixels identified as gray exceeds a prespecified threshold.[†] Whitespace filtering is performed by calculating the brightness of each pixel (average of red, green, and blue channels), classifying pixels with brightness above a given threshold[‡] as white, and discarding the tile as background if the proportion of pixels identified as white exceeds a prespecified threshold. Tile-level background filtering is performed at lower magnification for computational efficiency.

---

[*] Gaussian blur filtering by default is performed with σ=3 and threshold=0.02 at one-fourth the target magnification.

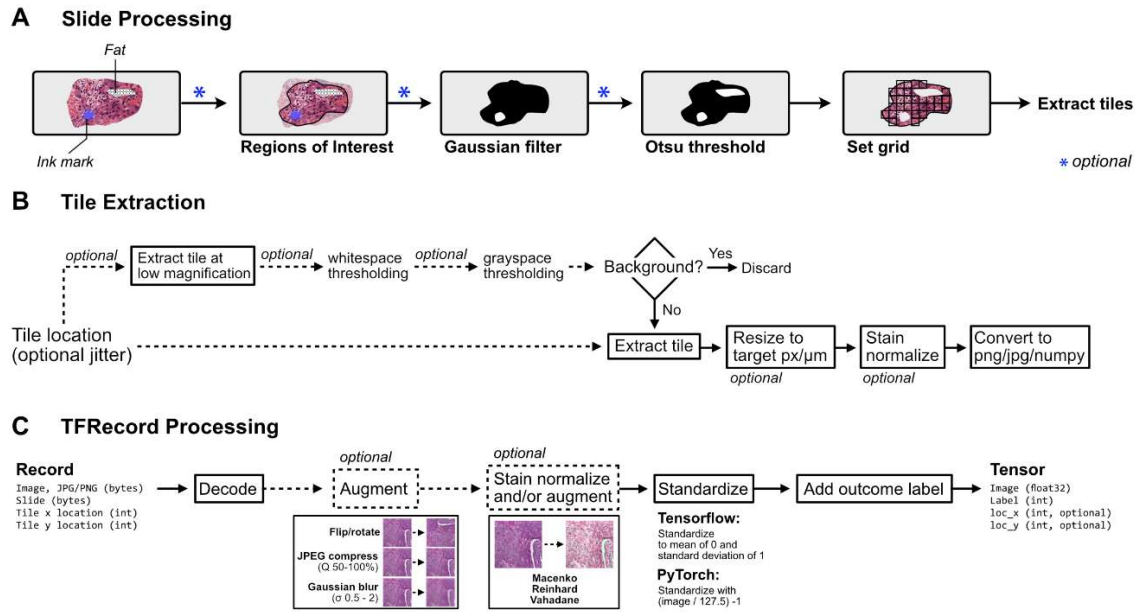

**Figure 1. Schematic of data flow during whole-slide image tile extraction and image processing.** (a) Schematic of initial slide processing and grid preparation. Whole-slide images can be annotated with Regions of Interest (ROI) to include only relevant areas of a slide for subsequent analysis. Optional slide filtering steps, including Gaussian blur filtering and Otsu's thresholding, may be applied at this step. The remaining areas of the slide are sectioned into a grid in preparation for tile extraction. (b) Data flow during tile extraction. If tile-level filtering is to be performed, such as whitespace or grayspace filtering, a low-magnification image at the highest pyramidal layer is taken at the given location and used for background filtering. If the tile passes filtering, the full-magnification image extracted, optionally resized to match a target micron size, stain normalized, and converted into a PNG image, JPEG image, or Numpy array. Extracted tiles can be saved to disk as individual files, or buffered into TFRecords for faster dataset reading. (c) Schematic of data flow when reading from TFRecords. Image tiles are buffered in TFRecords with JPEG or PNG compression and stored with slide and location metadata. During dataset iteration, data is decoded and converted to Tensors. Augmentation, including random flipping/rotation, random JPEG compression, and random Gaussian blur, can be applied at this step. If stain normalization was not performed during tile extraction, stain normalization can be applied at this step when iterating through a TFRecord dataset in real-time. Images are then standardized, and slide names are converted to ground-truth outcome labels.

The effective optical magnification at which tiles are extracted can be determined by either designating a pyramid level or a tile width in microns (**Fig. 2**). If a pyramid level is used, image tiles are extracted at a specified pixel size. Not all slides have the same optical magnifications available as pyramid layers, so this approach may require that some slides are skipped (**Supplementary Fig. 1**). Furthermore, this approach restricts the magnification levels that can be explored for downstream analysis to only what is available in each slide. Alternatively, if a tile width is specified in microns, image tiles will be extracted from the nearest higher-magnification pyramid layer and downsized to the target micron width, allowing researchers to granularly specify effective optical magnification. For this added flexibility, micron-based tile extraction is preferred. When using cuCIM, images are resized using bilinear interpolation, and when using VIPS, images are resized using Lanczos interpolation. Tile extraction is heavily optimized and multiprocessing accelerated, enabling whole-slide tile extraction at 40X magnification in as little as 2.5 seconds per slide. Fast tile extraction enables researchers to quickly experiment with different magnification levels.

---

[*] Default grayspace saturation filtering threshold is 0.05.

[†] Default grayspace fraction threshold is 0.6.

[‡] Default whitespace brightness filtering threshold is 230.

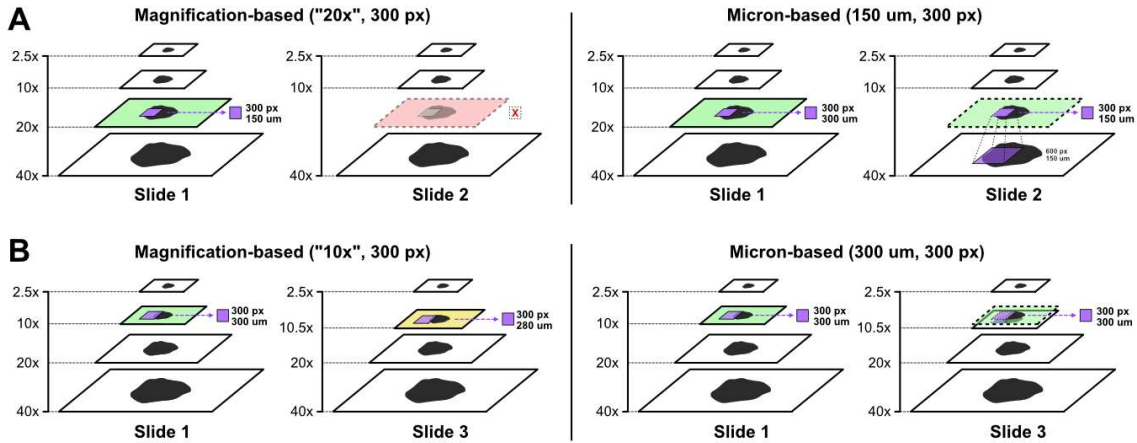

**Figure 2. Magnification-based vs. micron-based tile extraction.** (a) Comparison between magnification-based and micron-based tile extraction at 20x effective optical magnification. In this example, slide 1 has internal pyramid images stored at 2.5x, 10x, 20x and 40x magnification. Slide 2 has images stored at 2.5x, 10x, and 40x. Magnification-based strategies extract tiles at a matching layer in the image pyramid for a target optical magnification. In this example, Slide 2 is missing the 20x magnification layer, so tiles could not be extracted at 20x magnification. In comparison, with micron-based tile extraction, image tiles would be extracted at the 40x layer for Slide 2 and resized to an effective optical magnification of 20x. (b) Comparison between magnification-based and micron-based tile extraction at 10x effective optical magnification. In this scenario, the "10x" layers in Slide 1 and Slide 3 have slightly different effective optical magnifications – 10x and 10.5x – due to slide scanner differences. With magnification-based tile extraction, image tiles extracted from Slide 1 and Slide 3 would have slightly different effective optical magnification. With micron-based tile extraction, tiles would be extracted at the 10.5x magnification layer from Slide 3 and resized to match the same effective optical magnification as Slide 1 (10x). This strategy ensures that all image tiles have the same effective optical magnification.

All parameters needed to reproduce processed slide data are logged during tile extraction, assisting with data lineage tracking and management. Visual summary reports containing images of WSIs overlayed with tile masks and selected tiles from each WSI are automatically generated after tiles have been extracted, allowing researchers to quickly assess the quality of masking, filtering, and extracted images. This step can identify potential dataset issues, such as out-of-focus slides, suboptimal background or artifact removal, or low-quality ROIs. Slideflow Studio, an interactive whole-slide graphical user interface (GUI) included with Slideflow, can be used to preview slide masking, background filtering, and stain normalization settings in real-time. Interactive visualization assists with rapid determination of optimal slide processing parameters if further tuning of these settings is required. Slideflow Studio is described in more detail in section 2.11.

**2.3 Stain normalization and augmentation**

Digital hematoxylin and eosin (H&E) stain normalization, which is used to help reduce biasing batch effects incurred by differences in staining color and intensity among slides, can be applied to image tiles either during tile extraction or in real-time during model training. Available stain normalization methods include Reinhard [18], Macenko [19], and Vahadane [20], as well as a masked variant for Reinhard – where normalization is only applied in non-white areas – and fast variants for both Reinhard and Macenko normalizers, with the brightness standardization step disabled. The Reinhard and Macenko normalizers have native Numpy/OpenCV, Tensorflow, and PyTorch implementations to improve computational efficiency and enable GPU acceleration. Stain normalizers include several default reference fits and can be optionally fit to user-defined images.

Real-time stain augmentation can be performed during training when using the Reinhard or Macenko normalizers. Slideflow includes a novel stain augmentation approach performed by dynamically randomizing the stain normalization target, using a preset standard deviation around the normalization fit values. As the randomized stain matrix targets are centered around normalized values, the result is a combination of both normalization and augmentation. This approach differs from the stain augmentation described by Tellez *et al*, in which authors performed stain augmentation in the deconvoluted stain matrix space without normalization [21].

Additionally, Slideflow includes an option for using contextual information from the corresponding WSI during Macenko stain normalization. With contextual normalization, staining patterns across a slide are used to inform normalization of a constituent image tile. The deconvoluted H&E channels for a given image tile are normalized using the maximum H&E concentrations calculated from a context image, rather than calculating these maximum concentrations from the image being transformed. When this option is used, the context image is a thumbnail of the WSI, with background and areas outside ROIs removed to prevent pen marks and other artifacts from interfering with stain deconvolution. Contextual normalization is not recommended when artifact removal is suboptimal or ROIs are unavailable.

## 2.4 Training weakly-supervised, tile-based models

Slideflow includes several tools for training deep learning classification, regression, and time-to-event models for WSIs. Weakly-supervised, tile-based models use a deep learning image classifier to generate predictions for all tiles in a slide, and the final slide-level prediction is calculated by averaging tile predictions (**Fig. 3**). Although this method may be limited with highly heterogeneous tumors or when salient morphological features are sparse, it has proven to be an effective approach for a variety of biological applications [6], [22]–[24]. If tile-averaged predictions are insufficient for the research query, custom methods for slide-level prediction can be defined.

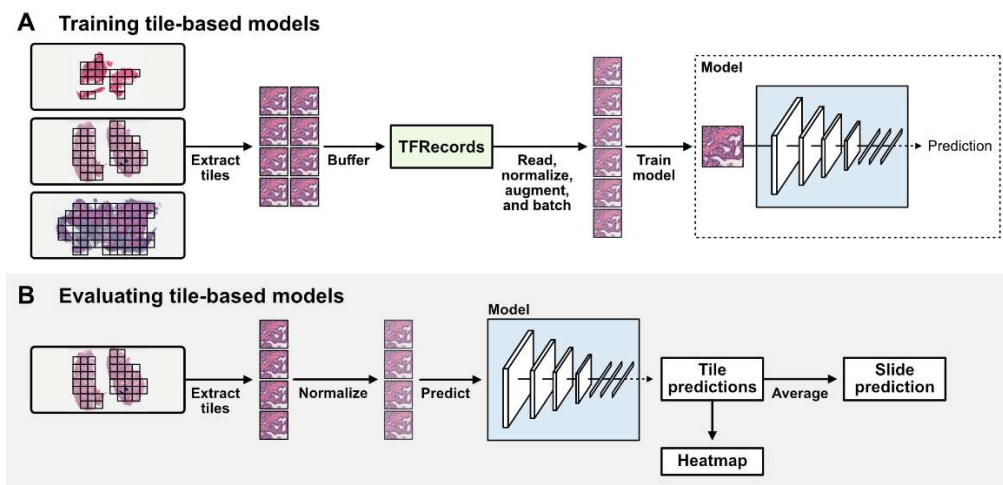

**Figure 3. Summary of approach for tile-based deep learning models.** (a) Image tiles are extracted from whole-slide images after any slide processing and buffered into TFRecords. During training, tiles are read from TFRecords, augmented, stain normalized, standardized, and batched. In this weakly-supervised approach, models are trained using ground-truth labels for each tile determined from the label of the corresponding slide. (b) When evaluating a tile-based model, image tiles do not need to be buffered into TFRecords. Image tiles can be extracted from slides, processed and stain normalized, and predictions are generated for each tile from a slide. The final slide-level prediction is the average prediction from all tiles. Tile-level predictions can be visualized as a heatmap.

Slideflow includes dataset organization tools to easily support the standard train, validate, test paradigm used for biomarker development. A held-out test set is first designated, either manually or using one of several tools for dataset splitting. Several approaches can be used for separating the remaining dataset into a training and validation splits, including fixed splitting, bootstrap, *k*-fold cross-validation, and site-preserved *k*-fold cross-validation [25].

Model architecture, loss function, and training hyperparameters are then configured; a list of included pre-configured model architectures is provided in **Supplementary Table 1**. Slideflow also supports custom models and loss functions using either Tensorflow or PyTorch. Models can be trained to single categorical, multi-categorical, continuous, or time-to-event outcomes. Multi-modal models can also be trained with additional arbitrary input, such as clinical variables, using late multimodal fusion via concatenation at the post-convolutional layer. Hyperparameters can be optionally tuned using either grid-search sweeps or Bayesian hyperparameter optimization, which uses the SMAC3 [26] package. Hyperparameter search spaces can be easily customized, and several preconfigured search spaces are provided.

During training, images can undergo augmentation to broaden the training domain and promote generalizability (**Fig. 1C**). Images can undergo random cardinal rotation and random horizontal or vertical flipping, random JPEG compression,[*] and random Gaussian blur,[†] and stain augmentation. When using Tensorflow, images are standardized to a mean of 0 and standard deviation of 1, and when using PyTorch, images are standardized to a range of 0-1.

By default, mini-batch balancing is used to ensure equal representation of all slides in each batch and equal representation of all classes (**Supplementary Fig. 2**). Throughout training, slides and classes will be oversampled to enable this balancing. This mini-batch balancing can be customized or disabled. During training, one epoch is defined as the total number of image tiles available in the training dataset divided by the batch size. The user can specify the interval at which validation checks are performed in the middle of an epoch. Early stopping can be configured to trigger when the exponential moving average of either loss or accuracy plateaus, or models can be trained for a prespecified number of epochs. Training can be distributed across multiple GPUs for computational efficiency. Training progress is monitored locally using Tensorboard [27] or remotely using Neptune.ai [28].

**2.5 Evaluating weakly-supervised tile-based models**

After training, performance metrics will be automatically calculated from the validation dataset at the tile-, slide- and patient-levels, including accuracy, area under receiver operator curve (AUROC), and average precision (AP). Slide-level predictions are calculated by averaging the one-hot predictions for all tiles from a slide, and patient-level predictions are calculated by averaging all tile-level predictions across all slides from a given patient. Tile-, slide-, and patient-level predictions are saved during both training and evaluation, allowing the researcher to calculate custom metrics if desired. Saved models can also be applied to held-out test sets, calculating test-set metrics, or to single slides for inference.

Predictive heatmaps are generated by overlying tile-level predictions onto a slide and can also be generated for either a single slide or multiple slides. Heatmap calculation is accelerated by parallelized tile extraction with multiprocessing. Heatmaps can be rendered and exported as images or interactively viewed with real-time navigation in Slideflow Studio, the whole-slide GUI.

**2.6 Uncertainty quantification**

Estimation of confidence and uncertainty is essential for medical AI applications, as reliable uncertainty quantification can facilitate clinical decision-making and potentially improve patient safety [29], [30].

---

[*] 10% chance of compression at a random quality level between 50-100%

[†] 50% chance of Gaussian blur with sigma between 0.5-2.0

Slideflow enables uncertainty estimation using the Monte Carlo dropout paradigm [31]. With this approach, models are built with dropout layers enabled during both training and inference. During inference, a single image undergoes multiple forward passes in the network, with the resulting distribution representing the final prediction (mean) and uncertainty (standard deviation). Tile-level uncertainty can be translated into slide-level uncertainty and used for subsequent confidence thresholding, using our previously described uncertainty thresholding algorithm [32]. Briefly, from a given set of validation predictions, an optimal uncertainty threshold is determined, below which predictions are more likely to be correct compared to predictions with higher uncertainty. Predictions with uncertainty above this threshold are then discarded. This thresholding is performed first at the tile level and then at the slide level. Thresholds are determined for a given training dataset using nested cross-validation, to prevent data leakage. This uncertainty estimation and confidence thresholding approach improves accuracy for high-confidence predictions and guards against domain shift.

**2.7 Image features and feature space analysis**

Converting images into feature vectors provides an avenue for feature space analysis and more advanced, whole-slide classification models such as multiple-instance learning (MIL). Three feature generation methods are provided for processing image tiles into numerical vectors: pretrained networks, finetuned classifiers, and self-supervised learning (**Fig. 4**). For all methods, features can be calculated for single image tiles, a single slide, or image tiles read from TFRecords.

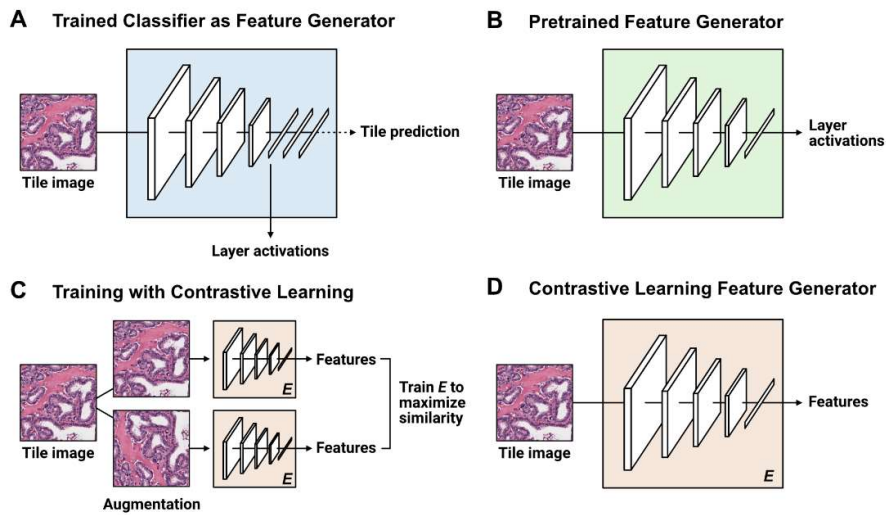

**Figure 4. Feature generation methods.** (a) Trained classifiers can be converted into feature generators by specifying a neural network layer and calculating activations at the given layer. (b) Several pretrained models can be used as feature generators, including non-pathology pretrained models (*e.g.* ImageNet), or pathology-specific pretrained models (*e.g.* CTransPath). (c) Self-supervised contrastive learning (SimCLR) can be used to train a feature generator without requiring ground-truth labels for classification. (d) Features can be calculated from a model trained using self-supervised learning.

Several pretrained networks can be used for converting images into feature vectors. Slideflow includes an API for calculating layer activations at any arbitrary neural network layer for models pretrained on ImageNet, by specifying an architecture name and layer name. If multiple layers are specified, activations at each layer will be calculated and concatenated. An API is also included for easily generating features using the pretrained, pathology-specific CTransPath [33] and RetCCL [34] networks. Finetuned classifiers trained with Slideflow can also be used for feature generation, calculating activations at any arbitrary neural network layer. Finally, an API is included for easily training the self-supervised learning model SimCLR, providing another avenue for feature generation through contrastive pretraining.

Once features are calculated for a dataset of image tiles, several tools are provided for flexible dimensionality reduction and feature space visualization using UMAP [35]. UMAP plots can be quickly generated and labeled with outcomes, clinical variables, and suspected confounders such as site. Corresponding image tiles can be overlaid onto the UMAP plots to aid in interpretability, resulting in visualizations we refer to as mosaic maps. Mosaic maps are generated by separating a given UMAP projection into a grid (default 50x50), and for each grid space, plotting the corresponding image tile for any of the points in the grid space. Tile selection within a grid space can either be random (default), or the tile nearest to the centroid for that grid space can be chosen. Mosaic maps can be viewed alongside labeled UMAPs, providing insights into the relationship between morphologic image features, the outcome of interest, and other relevant clinical factors. This process can also help identify potential sources of confounding and bias. Mosaic maps can be exported as a high-resolution figure or viewed interactively with Slideflow Studio.

### 2.8 Weakly-supervised multiple-instance learning

Tile-based whole-slide classification models may not be well-suited for datasets where relevant histopathological features are expected to be sparse or heterogenous in a slide, or when pathologist-annotated regions of interest are unavailable. Attention-based multiple-instance learning (MIL) models provide an avenue for weakly-supervised whole-slide classification that is theoretically still robust when relevant features are sparse or regions of interest are unavailable [36]. These models generate predictions from bags of image tile feature vectors, and models will learn to ignore uninformative image tiles through attention weighting. The performance of these models is in part dependent upon the quality of features calculated from image tiles.

Three types of MIL models can be trained in Slideflow, including traditional MIL [37], attention-based MIL [38], and clustering-constrained attention-based MIL (CLAM) [39]. An API is provided for fast conversion of image tiles into feature vectors and training of MIL models from these generated features. Training is executed using the FastAI framework [40], including options for either cosine annealing learning rate scheduling or one-cycle learning rate scheduling [41]. For CLAM, an option is provided to use the originally published training loop instead of the FastAI trainer, if desired. Trained MIL models can be used to generate attention heatmaps for WSIs, highlighting areas of the slide weighted with high attention. As with predictive heatmaps for tile-based classification models, attention heatmaps can be exported as high-resolution figures or interactively viewed in Slideflow Studio.

### 2.9 Generative adversarial networks

A growing body of evidence is showing the potential utility of Generative Adversarial Networks (GANs) for histopathological applications. GANs can reproduce realistic synthetic histology images that have been used for training augmentation, stain and color normalization, image enhancement, and explainability. Slideflow includes an API for training and using StyleGAN2 [42] and StyleGAN3 [43] with optional class conditioning from image tiles saved in TFRecords. Model architectures, training paradigms, and configuration options are all equivalent to their original implementations. The API provides a method for training these GANs from preprocessed images already stored in TFRecords, without requiring additional data processing or alternative formatting, and additionally provides an interface for calculating predictions from generated images using a trained classifier.

For class-conditional GANs, several tools are provided for generating images from intermediate classes using feature space embedding interpolation. To generate an image in between two classes, we calculate the associated class embedding for each class, and then perform a linear interpolation to achieve an intermediate embedding. These intermediate embeddings can be used for class or layer blending applications [44]. Images generated from trained GANs can be exported as raw PNG or JPG images, saved in TFRecord format, or visualized in real-time using Slideflow Studio.

**2.10 Model explainability**

Explainable artificial intelligence (XAI) approaches are an increasingly important component of medical imaging research. These techniques can provide insights into what image features models have learned, support biological plausibility of model predictions, and improve model trust [13]. Slideflow includes four methods for model explainability: heatmaps, mosaic maps, gradient-based pixel attribution (saliency maps), and conditional generative adversarial networks (cGANs).

Heatmaps, including both predictive and attention heatmaps, provide an avenue for quickly assessing the areas of a WSI relevant to the final prediction. Described above, heatmaps can be calculated for a single slide or a dataset of multiple slides, and either exported as a high-resolution figure or interactively viewed.

Mosaic maps provide a tool for feature space exploration at arbitrary neural network layers. With pathologist interpretation, they can highlight morphologic associations between outcome variables and offer insights into the spatial relationship of image features among classes.

Gradient-based pixel attribution approaches highlight the pixels in an image that were relevant for a given neural network model prediction. These attribution heatmaps, or saliency maps, can be calculated through a variety of provided methods, including Grad-CAM [45], vanilla gradients [46], integrated gradients (and variations thereof) [47], and XRAI [48]. Saliency maps can be displayed as raw heatmaps or as overlays onto the associated image. Utility functions are included for rapid comparison between different saliency map methods. Saliency maps can also be interactively viewed for WSIs in real-time using Slideflow Studio.

cGANs offer a dataset-level explainability approach for trained neural networks models, using the paradigm we have recently described [44]. The synthetic histology generated by cGANs can illustrate morphologic features associated with outcome classes and can provide insights into the importance of larger histopathological features not amenable to localization with saliency maps, such as differences in architecture, stroma, colloid, and staining patterns.

**2.11 Whole-slide visualization with Slideflow Studio**

Slideflow includes a visualization tool, Slideflow Studio, for interactively viewing WSIs, focal predictions, heatmaps, saliency, uncertainty, and mosaic maps. Slide imaging data is read using either cuCIM or VIPS, accelerated with multiprocessing, and rendered using OpenGL. Slide processing settings, such as Otsu's thresholding, grayspace filtering, stain normalization, *etc*., can be previewed in real-time to assist with fast determination of optimal slide processing parameters (**Supplementary Fig. 3**). ROI annotations can be loaded, edited, and added with a lasso selection tool.

Once a model is loaded, right-clicking anywhere on the slide will reveal a preview window showing an image tile extracted at this location, before and after stain normalization (if applicable) (**Fig. 5**). Saliency maps generated through gradient-based pixel attribution can also be displayed as a heatmap or as an overlay on the extracted image tile. Tile-level predictions and uncertainty are shown in the control panel.

Whole-slide predictive heatmaps can be calculated, customized, and displayed for slides once a model is loaded. A "low memory" mode can be enabled in Performance Settings, reducing memory consumption at the cost of slower heatmap calculation. Final slide-level predictions (and uncertainty, if applicable) will be shown in the control panel after the heatmap is calculated.

Mosaic maps can also be interactively viewed in Slideflow Studio (**Supplementary Fig. 4**). Compared with a static figure, this interface loads images faster, permits closer inspection of image tiles at higher resolution through zoom, and enables dynamic modification of grid resolution to increase or decrease the granularity of the mosaic grid. A popup window in the bottom-right corner shows the user the corresponding UMAP plot, with a red box indicating the current section of the plot in view. Hovering over an image tile will show a larger section from the corresponding slide at that tile location, revealing the surrounding histologic context.

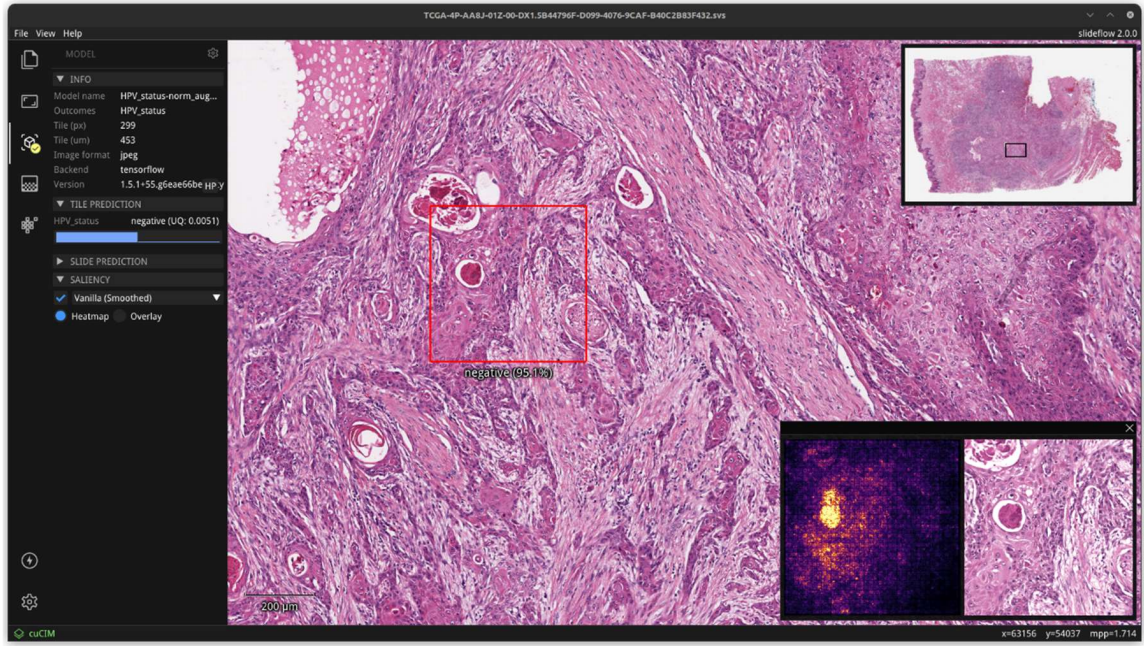

**Figure 5. Real-time predictions and saliency maps from whole-slide images.** Slideflow Studio provides an interface for generating both whole-slide and focal tile predictions. Right clicking on an area of the screen extracts a tile at the given location, showing the tile before and after stain normalization (if applicable). Gradient-based pixel attribution methods can be used to render saliency maps for the given image, which can be shown as a heatmap (shown here) or as an overlay (not shown).

## 2.12 Supported hardware and software environments

Slideflow is a Python package that requires Python 3.7 or greater, and either Tensorflow or PyTorch. Model training requires a Linux-based operating system, such as Ubuntu or RHEL/CentOS, with a dedicated NVIDIA GPU. Trained models can be deployed using the interactive interface on Linux-based systems, Windows, macOS (Intel and Apple chip), and ARM-based devices such as the Raspberry Pi.

# 3 Results

The variety of tools offered by Slideflow have been utilized for many diverse research objectives, including gene expression prediction [28], prognostication [49], uncertainty quantification [32], identification of bias and batch effects [25], drug response prediction [50], and generation of synthetic histology for model explainability [44]. In order to illustrate some of these tools on a real dataset, we will present results on a benchmark dataset for Human Papilloma Virus (HPV) status prediction in head and neck squamous cell carcinoma. The training/validation dataset is comprised of 262 patients (151 HPV-negative, 111 HPV-positive) from the University of Chicago, and the held-out external dataset is comprised of 459 patients (407 HPV-negative, 52 HPV-positive) across 26 sites from The Cancer Genome Atlas (TCGA). All patients had one associated WSI.

## 3.1 Slide processing

Pathologists annotated ROIs encircling areas of tumor for all slides, except in cases where the entire sample was determined to be tumor. Otsu and Gaussian blur filtering were explored for this dataset. In some cases, Otsu's thresholding highlighted pen marks as foreground tissue, so both background filtering methods were used in addition to ROIs for all slides. Image tiles were extracted at 299 pixels and seven micron sizes ranging from 76 μm (40X magnification) to 1208 μm (2.5X). Using the cuCIM backend, tile extraction speed ranged from 614 – 2091 tiles/second (0.38 – 25 slides/second) (**Fig. 6, A and B**). Tile extraction was also performed using VIPS for comparison, with tile extraction speed ranging between 198 – 1300 tiles per second (0.24 – 11.5 slides/second). Otsu's thresholding added 0.30 ± 0.12 seconds per slide, and Gaussian blur filtering added 1.39 ± 0.67 seconds per slide. Tile extraction with cuCIM was 1.6 - 3.3 times faster than VIPS (**Fig. 6, A and B**). Example pages from the associated tile extraction PDF reports are shown in **Supplementary Figs. 5 and 6**. TFRecord buffering permitted dataset iteration at 11,453 images/second using Tensorflow and 6,784 images/second using PyTorch. Buffered dataset sizes are shown in **Fig 6C**.

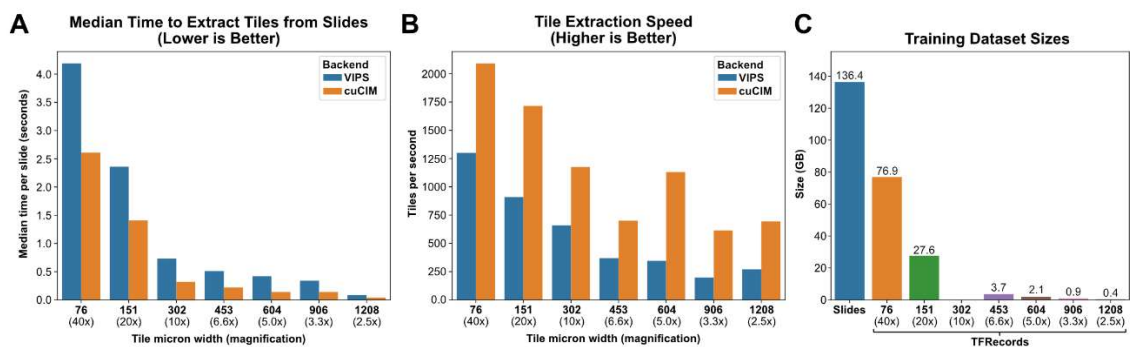

**Figure 6. Whole-slide image processing speed and dataset sizes.** (a) Median time required to extract tiles from WSIs using each of the two available slide processing backends. (b) Average tile extraction speed from WSIs, using each slide processing backend. (c) Size of the full training dataset as raw slides and in buffered TFRecords, at each assessed magnification size. Image tiles were stored in TFRecords using JPEG compression with 100% quality. Benchmarks were performed using an AMD Threadripper 3960X.

Experimentation with tile-based background filtering was performed for comparison but not used for downstream analysis (**Fig. 7**). Otsu's thresholding identified 6.4% ± 8.2% more background tiles than grayspace filtering. Image tiles removed by Otsu's thresholding but not grayspace filtering typically included edge tiles or images with heterogenous background content, such as fat or stroma (**Fig. 7B**). With rare exception, all image tiles removed by Otsu's thresholding were also removed with grayspace filtering. For 85.2% of slides, there was a <5% difference in background identified by whitespace filtering and grayspace filtering. In 12.6% of cases, grayspace filtering removed a median of 10.2% more background than whitespace filtering (**Fig. 7C**), and in 2.2% of cases, whitespace filtering failed to remove any background tiles.

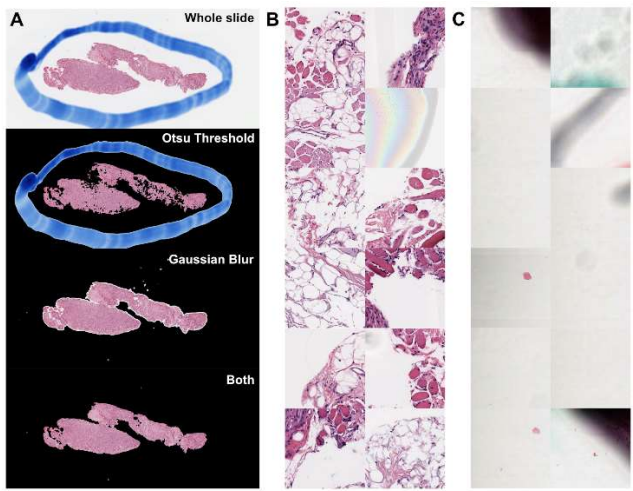

**Figure 7. Comparison between background filtering methods.** (a) Comparison between slide background filtering methods. An example whole-slide image with large pen mark artifact is shown, with associated background filtering masks applied. Black areas indicate masked background. With Otsu's thresholding alone, the pen mark is identified as foreground, and several parts of the tissue area are erroneously removed as background. Gaussian blur filtering removes the pen mark, but some smaller areas of background near edges are not removed. Performing Gaussian blur first, followed by Otsu's thresholding, results in the most accurate background identification and removes the pen mark. (b) Example images identified as background using Otsu's thresholding, but not when using grayspace filtering. (c) Example images identified as background using grayspace filtering, but not when using whitespace filtering.

Three stain normalization methods with several variations were compared, as shown in **Fig. 8**. The fast variants of the Reinhard methods produced similar results to the standard Reinhard variants, with slight differences in perceived brightness. The masked and unmasked Reinhard variants yielded similar images in most contexts, but with the unmasked variants producing pink-tinted background for image tiles containing high background content. Compared with standard Macenko normalization, context-aware Macenko normalization generally resulted in images with higher perceived contrast but occasionally washed out fine details in bright areas, such as areas containing fat (**Supplementary Fig. 7**). Macenko stain augmentation yielded realistic, artifact-free images with diverse staining hues (**Supplementary Fig. 8**).

Computational efficiency was compared between methods using both Tensorflow and PyTorch, with results shown in **Fig. 9**. Removal of the brightness standardization step improved peak stain normalization speed by 272% for the Reinhard normalizer and 51% for the Macenko normalizer when using Tensorflow, and 26% and 24% for Reinhard and Macenko normalizers when using PyTorch. Reinhard normalizers exhibited superior computational performance when processed on the GPU, and Macenko normalizers were faster when processed on CPU. The utility of the Vahadane algorithm is limited by long processing times, and could not be used for real-time normalization. Based on a combination of qualitative assessment and computational efficiency, the standard Macenko stain normalization strategy was chosen for subsequent analyses. Real-time stain augmentation was used during model training.

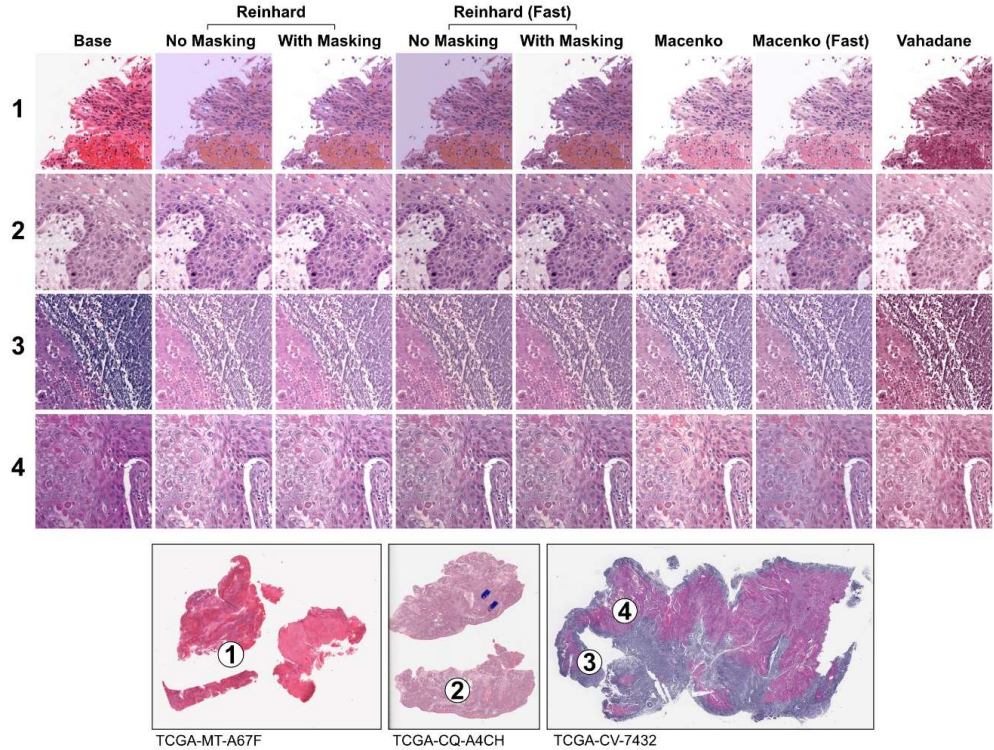

**Figure 8. Comparison between stain normalization methods**. Four image tiles from whole-slide images with different staining patterns are shown before and after stain normalization. Three stain normalizers are compared: Reinhard, Macenko, and Vahadane. The Reinhard normalizer has a standard and fast variant (with brightness standardization disabled), and a masked and unmasked variant. Similarly, the Macenko also has a standard and fast variant.

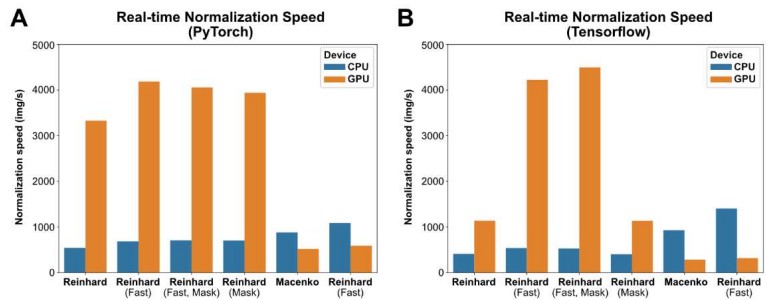

**Figure 9. Real-time stain normalization benchmarks.** Stain normalization speed was assessed for each normalization method and device type (CPU, GPU) using both Tensorflow and PyTorch. (a) Benchmark results using PyTorch. (c) Benchmark results using Tensorflow. All benchmarks were obtained using an AMD Threadripper 3960X CPU and A100 40 GB GPU.

### 3.2 Weakly-supervised tile-based classification

Weakly-supervised, binary classification models were trained on the institutional dataset of 262 slides to predict HPV status using three-fold cross-validation and the Tensorflow backend. As a first step, models were trained on the first cross-fold at seven magnification levels between 2.5X and 40X using the Xception architecture and a single set of hyperparameters (**Fig. 10**). The best performance was seen at 6.6X (tile micron width of 453), so this magnification level was used for subsequent analysis. Hyperparameters were tuned on the first cross-fold with Bayesian hyperparameter optimization using the "shallow" search space configuration, a maximum of 50 iterations, and 5 replicate models trained for each hyperparameter combination. Average training time for each model was 2 minutes 11 seconds. Ten models were trained using the best performing hyperparameter combination and compared with ten models trained using the starting hyperparameters. AUROC was not improved using the optimized hyperparameters (0.808 vs. 0.803, P=0.11), so the initial hyperparameters were used for subsequent analysis (**Supplementary Table 2**). AUROC across the three cross-folds was 0.80, 0.84, and 0.78, with AP of 0.79, 0.83, and 0.77 (**Fig. 11**). Dropout-based uncertainty quantification was used for confidence threshold determination. A final model was then trained across the full dataset using the previously determined optimal hyperparameters and the Tensorflow backend. When validated on the external test set comprised of 459 slides from TCGA, the model resulted in an AUROC of 0.87 and AP of 0.80. Using uncertainty quantification and confidence thresholding, 89.5% of slides had high-confidence predictions, with an AUROC of 0.88 within high-confidence predictions. Using prespecified prediction and uncertainty thresholds determined from cross-validation, the final model had a test-set accuracy of 90.2%, sensitivity of 77.0%, and specificity of 92.0% within high-confidence predictions.

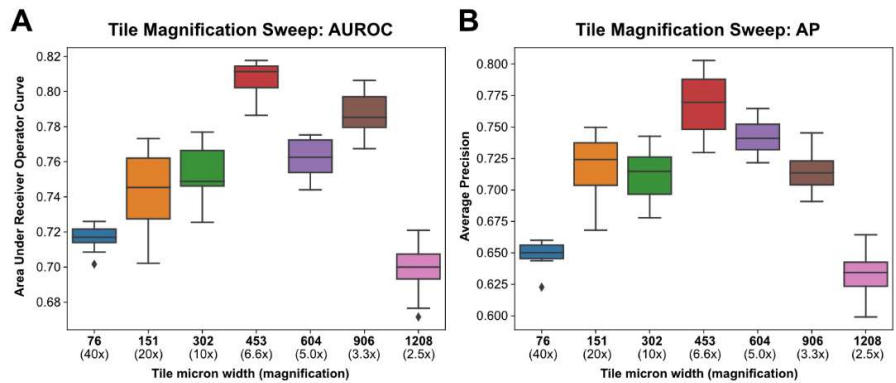

**Figure 10. Tile magnification optimization.** Image tiles were extracted from slides in the training dataset at seven magnification sizes, ranging from 2.5x (tile width 1208 microns) to 40x (tile width 76 microns). Using the first training cross-fold, ten replicate models were trained at each magnification size using random weight initialization. (a) Area Under Receiver Operator Curve (AUROC) for models trained at varying magnifications. (b) Average Precision (AP) for models trained at varying magnifications.

After the final model was trained and evaluated on the external test set, a post-hoc analysis was performed to assess the impact of stain augmentation on model performance. Ten replicate models were trained with and without stain augmentation on the full University of Chicago training dataset and then evaluated on the TCGA external test set. This post-hoc analysis demonstrated that stain augmentation resulted in a small but statistically significant improvement in AUROC on the test set, from $0.871 \pm 0.008$ to $0.882 \pm 0.014$ (P=0.021, one-sided *t*-test).

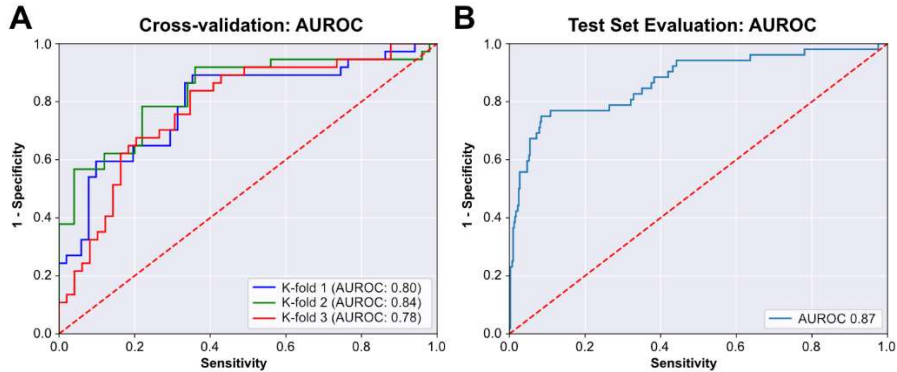

**Figure 11. Area Under Receiver Operator Curve (AUROC) for cross-validation and the held-out test set.** (a) AUROC for the three models trained in cross-validation on the single institution, University of Chicago dataset. (b) After cross-validation, a final model was trained on the full University of Chicago dataset. This model was then evaluated on the held-out test set from The Cancer Genome Atlas (TCGA). AUROC for this test set evaluation is shown.

### 3.3 Multiple-instance learning

In order to train multiple-instance learning (MIL) models, image features were first calculated for image tiles extracted at 6.6X magnification (453 μm width) using the pretrained model CTransPath [33]. Image tiles were normalized with Macenko normalization prior to feature calculation. MIL models were built using the CLAM single-branch architecture [39] and trained using a single set of default hyperparameters (**Supplementary Table 3**). Models were trained for a total of 20 epochs. Three-fold cross-validation AUROC for the MIL models was 0.77, 0.78, and 0.79. A final model was trained on the full University of Chicago dataset without validation, and then tested on the TCGA test set. On this held-out test set, the final model had an AUROC of 0.81 and AP of 0.77.

### 3.4 Feature space analysis

Features for the weakly supervised tile-based model were generated and visualized for both datasets using the included feature generation interface. Features are generated through calculation of post-convolutional layer activations for all image tiles, and the resulting feature space is visualized through UMAP dimensionality reduction. UMAP plots of the TCGA and University of Chicago feature spaces show good class separation between HPV-negative and HPV-positive images for both the training and test data (**Figure 12, A and B**). The mosaic map for the University of Chicago feature space highlights known biologically-relevant image features associated with HPV status **(Figure 12E)**. Area 1, enriched for HPV-positive images, shows image tiles with tightly packed cells with scant cytoplasm and surrounding inflammation. Area 3, enriched primarily with HPV-negative images, shows heavy keratinization along with pleomorphic cells with increased cytoplasm. Findings in both of these areas are consistent with known histopathological associations [51], [52]. Area 2, an intermediate zone with both HPV-positive and HPV-negative images, shows image tiles with keratinization, inflammation, and cells with varying cytoplasmic content. Together, this feature space analysis supports the biological plausibility of model predictions.

Features from CTransPath, which were used for training the multiple-instance learning model, were generated and visualized using the included feature generation interface. Separation between HPV-negative and HPV-positive image tiles on the University of Chicago training dataset using CTransPath features is less clear than when using features from the tile-based model (**Figure 12C**). On the test dataset from TCGA, there is no clear separation between HPV-positive and HPV-negative image tiles with this visualization (**Figure 12D**). This poorer separation between HPV-positive and HPV-negative images in the feature space may help partially explain the discrepancy in performance between the tile-based and MIL models.

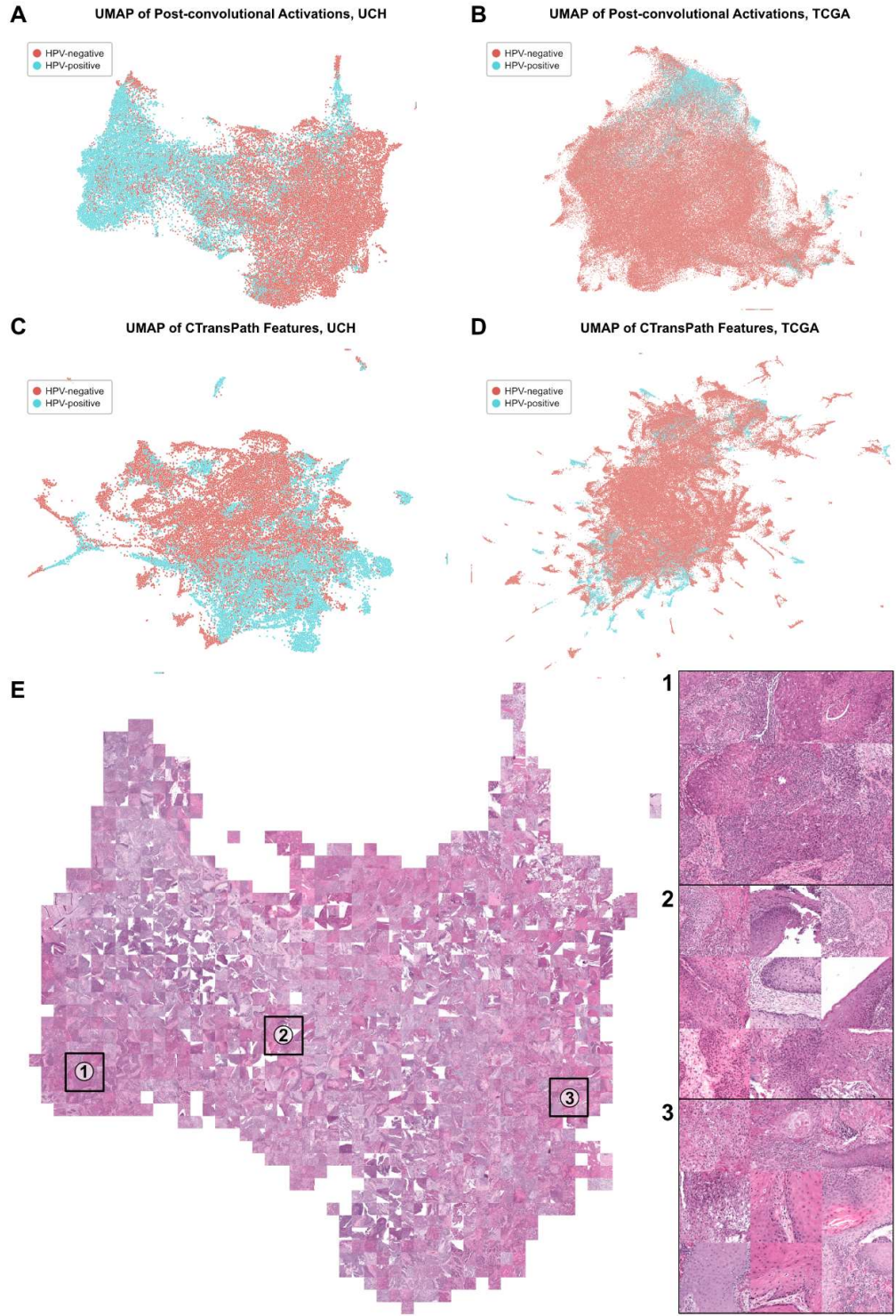

**Figure 12. Feature space visualization for the tile-based and multiple-instance learning models.** (a) UMAP plot of post-convolutional layer activations, calculated using the final trained model, for all images in the University of Chicago training dataset. (b) UMAP plot of post-convolutional layer activations, calculated using the final trained model, for all images in the TCGA test set. (c) UMAP plot of CTransPath features for all images in the University of Chicago training dataset. (d) UMAP plot of CTransPath features for all images in the TCGA test set. (e) Mosaic map generated from the UMAP plot shown in (a). Three areas are magnified for closer inspection. Area 1 is enriched for HPV-positive images, Area 2 is in a zone of transition between HPV-positive and HPV-negative images, and Area 3 is enriched with HPV-negative images. Image tiles are shown using Macenko stain normalization.

## 3.5 Explainability

Heatmaps of model predictions were generated from the final tile-based model at an average rate of 22.1 ± 4.2 seconds per slide. Example heatmaps of predictions and uncertainty are shown in **Figure 13**. In general, areas with low uncertainty and strong predictions for negative HPV status demonstrated high keratinization and cells with pleomorphic nuclei. Low-uncertainty areas with strongly positive HPV status predictions tended to show tightly packed cells with monotonous nuclei and surrounding inflammatory infiltrate. Both of these observations are consistent with known histopathological associations with HPV status [51], [52]. Attention heatmaps were also generated using the final MIL model, as shown in **Supplementary Figure 9**. Areas with strong, high-confidence predictions from the tile-based model were generally also weighted with high attention from the MIL model. A screenshot of the whole-slide user interface with a loaded heatmap is shown in **Figure 14**.

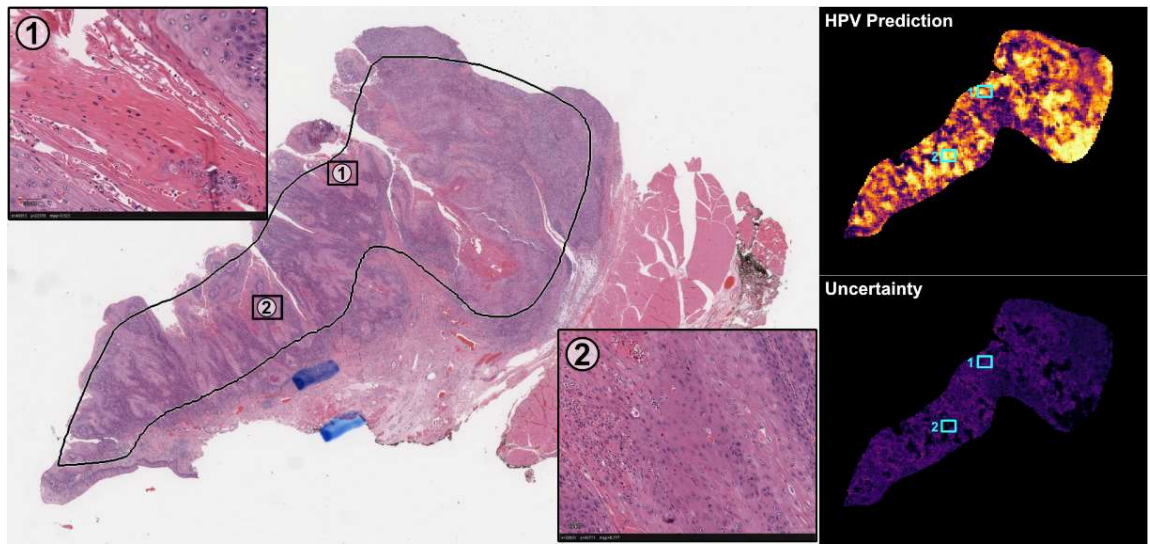

**Figure 13. Prediction and uncertainty heatmaps on an example slide in the external test set.** Heatmaps of HPV predictions and uncertainty were generated from the final tile-based model for a randomly selected slide from the test set. Two areas are shown with magnified display. Area 1 is a location with intermediate predictions and high uncertainty, showing mostly stroma and out-of-focus cells in the top-right corner. Area 2 is a location with strong HPV-positive predictions and low uncertainty, showing heavy keratinization and pleomorphic nuclei.

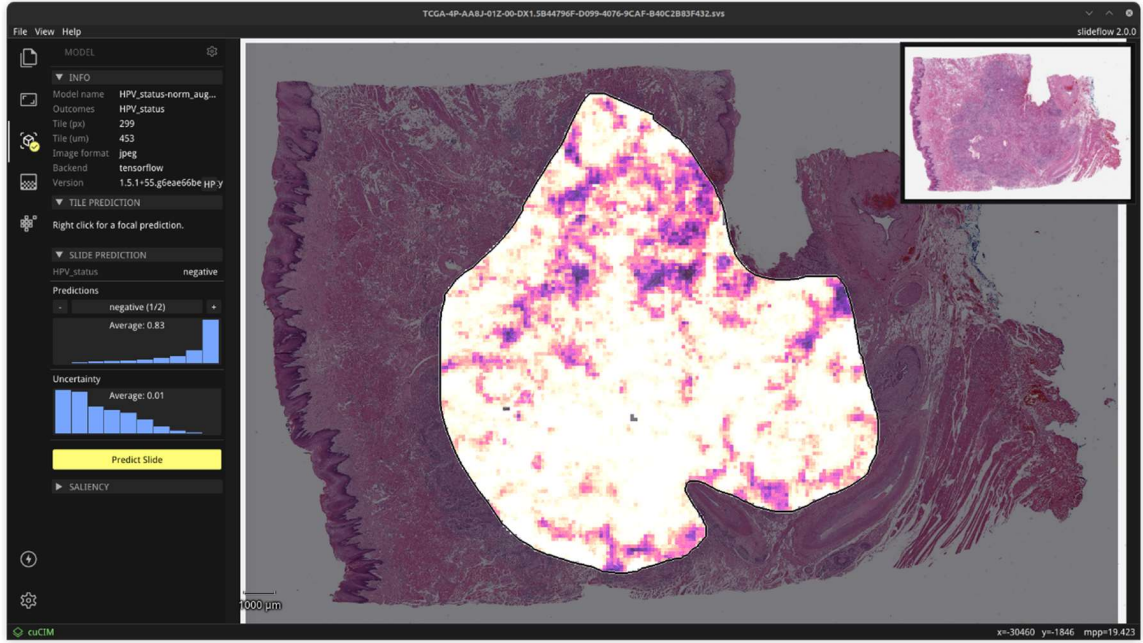

**Figure 14. Interface for viewing and navigating whole-slide heatmaps.** Slideflow Studio includes an interface for generating whole-slide predictions and heatmaps, as shown in this figure. Heatmaps can be viewed interactively and exported as both PNG images and Numpy arrays. Heatmap color and display options are customized in another tab of the interface (not shown).

Saliency maps were generated for an example correctly predicted HPV-negative image tile using vanilla gradients, three variations on integrated gradients, and XRAI (**Figure 15A**). For this example, all saliency maps highlight the section of the image tile with heavy keratinization, a histopathological factor known to be associated with HPV negativity. Finally, a conditional generative adversarial network (cGAN) was trained on the institutional dataset, conditioned on HPV status, to provide generative explanations for the trained classifier through synthetic histology [44]. The StyleGAN2 architecture was used, using default hyperparameters from the original implementation [42]. Training was stopped after 15 million images due to divergence with further training. Training took 38 hours on four A100 GPUs. Visualizations generated with this method highlighted differences in keratinization and nuclear pleomorphism, which is increased in the synthetic HPV-negative images, and inflammatory infiltrate, increased in synthetic HPV-positive images (**Figure 15B**). These differences are consistent with known pathologic associations with HPV status, further supporting the biological plausibility of learned image features [51], [52].

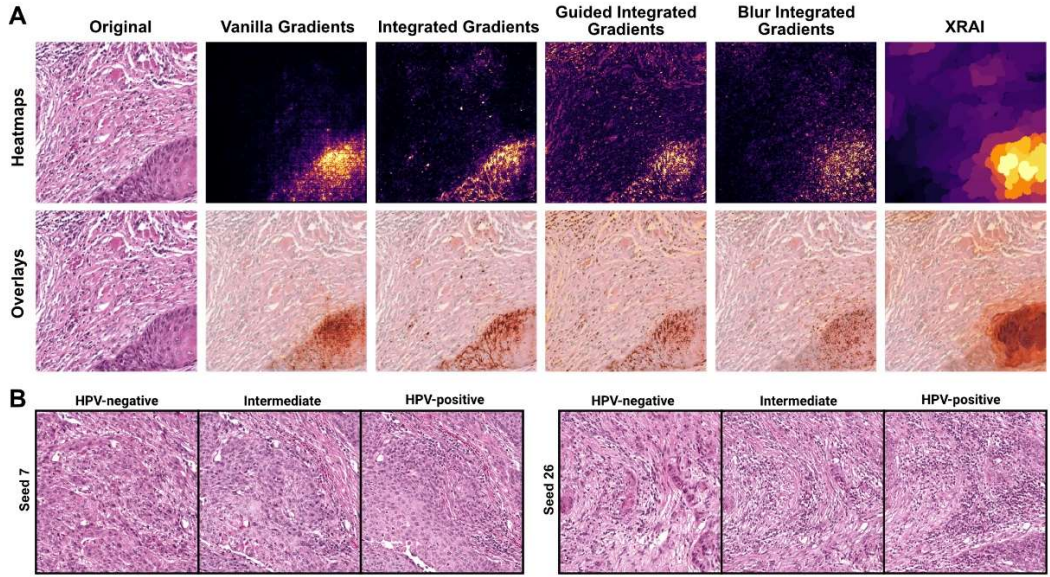

**Figure 15. Example of model explanations using saliency maps and generative adversarial networks. (a)** Various gradient-based pixel attribution methods with APIs available in Slideflow were used to generate saliency maps for an example HPV-negative image. **(b)** A class-conditional GAN based on the StyleGAN2 architecture was trained using Slideflow to generate synthetic histology images belonging to HPV-negative and HPV-positive negative classes, visually highlighting morphologic differences between classes.

### 3.6 Hardware deployment

The above experiments and benchmarks were performed on a Linux-based workstation with a dedicated NVIDIA GPU. To assess the feasibility of using the whole-slide graphical interface as a deployable tool for WSI analysis, Slideflow Studio was also deployed and tested on a Windows 10 desktop (with dedicated GPU), an Intel MacBook Pro, an M2 MacBook Pro, and a Raspberry Pi 4 (4 GB). All devices ran Slideflow Studio with usable performance and successfully generated predictions for WSIs. GPU acceleration and model training is only available on systems with a dedicated NVIDIA GPU.

## 4 Conclusions

Slideflow is a flexible, end-to-end deep learning toolkit for digital pathology with computationally efficient whole-slide image processing, data storage cross-compatible with Tensorflow and PyTorch, efficient GPU-accelerated stain normalization, and a GUI to support deployment of trained deep learning models. Slideflow includes a variety of digital pathology deep learning methods to support a wide range of research objectives without switching software environments or reprocessing data. The software is well documented and has been built to support researchers with a range of programming experience in the development of novel deep learning applications for digital histopathology.

## 5 Availability

Slideflow and Slideflow Studio are available at https://github.com/jamesdolezal/slideflow, via the Python Package Index (PyPI), and Docker Hub (https://hub.docker.com/r/jamesdolezal/slideflow). Slideflow is licensed with GNU General Public License v3.0.